\documentclass[twocolumn,showpacs,superscriptaddress,preprintnumbers,floatfix]{revtex4}
\usepackage{latexsym}
    \def\be{\begin{equation}}
    \def\ee{\end{equation}}
    \def\ba{\begin{eqnarray}}
    \def\ea{\end{eqnarray}}

\begin{document}
\title{Adiabatic regularization of the graviton stress-energy tensor in 
de Sitter space-time}

\author{F. Finelli} \email{finelli@bo.iasf.cnr.it}
\affiliation{CNR-INAF/IASF, Istituto di Astrofisica Spaziale e 
    Fisica Cosmica \\ Sezione di Bologna \\ Via Gobetti 101, 
I-40129 Bologna -- Italy}

\author{G. Marozzi} \email{marozzi@bo.infn.it}
\author{G. P. Vacca} \email{vacca@bo.infn.it}
\author{G. Venturi} \email{armitage@bo.infn.it}
\affiliation{Dipartimento di Fisica, Universit\`a degli Studi di Bologna
    and I.N.F.N., \\ via Irnerio, 46 -- I-40126 Bologna -- Italy}
\begin{abstract}
We study the renormalized energy-momentum tensor of gravitons in a 
de Sitter space-time. After canonically quantizing only the physical 
degrees of freedom, we adopt the 
standard adiabatic subtraction used for massless minimally coupled scalar 
fields as a regularization 
procedure and find that the energy density of gravitons in the 
$E(3)$ invariant vacuum is proportional to $H^4$, 
where $H$ is the Hubble parameter, but with a positive sign.
According to this result the scalar expansion rate, which is gauge
invariant in de Sitter space-time, is increased by the fluctuations.
This implies that gravitons may then add to conformally coupled matter in 
driving the Starobinsky model of inflation.

\end{abstract}

\maketitle

Interest in the back-reaction of particles produced by an external
field has always been great in the last thirty years. This issue has been 
studied in cosmological 
space-times following the seminal 
works by L. Parker \cite{parker}.
The back-reaction problem for quantized test
fields of different spin has been examined in great detail over
the decades for a de Sitter space-time \cite{BDbook}. 

Gravitational back-reaction treated in a self-consistent way is much 
more difficult. Taking into account metric fluctuations together with 
matter fluctuations to first order is a textbook subject. To second 
order, the level of difficulty increases due to non-linearity and to the 
invariance under coordinate transformations. 

Vacuum space-times are the simplest arena in which to study the energy 
content carried by metric fluctuations, since the latter just 
reduce to gravitational waves (i.e. tensor modes) 
in the absence of dynamical matter.
Even in such a simple setting, non-linear self-consistent solutions  
are quite surprising: the gravitational geon \cite{geon} is one of the 
historic examples, which triggered more work in this area.

Although gravitational waves are somewhat similar to massless minimally 
coupled test scalar fields, they are not exactly the same and the former 
may reveal quantum gravity aspects which are not proper of the latter: 
here we shall show this relevant difference in the de Sitter case.
In this paper we study the energy-momentum tensor (EMT henceforth) of 
quantized gravitons and the scalar expansion rate in de Sitter space-time 
in a perturbative way. We shall consider only half of the de Sitter 
Carter-Penrose diagram, i. e. only the cosmological flat expanding branch.
To our knowledge this case has been only 
approached by Ford \cite{ford}, but not fully explored. 
Since inflation is very close to de Sitter, this calculation may 
also be useful in the context of back-reaction during 
inflation. For the case of inflation initiated by quantum 
anomalies of test fields \cite{starobinsky}, this would lead to the 
complete evaluation of the gravitational sector.
Throughout units are chosen such that $\hbar = c =1$.

According to the action
\be
S = \frac{1}{16 \pi G} \int d^4 x \sqrt{-g} \left[ R - 2 
\Lambda \right]
\ee
the Einstein equations are:
\be
R_{\mu \nu} - \frac{1}{2} g_{\mu \nu} R + \Lambda g_{\mu \nu} = 0 
\ee

In the de Sitter space-time the only non-vanishing metric 
fluctuations are the dynamical degrees of freedom 
(the gravitons): 
\begin{eqnarray}
ds^2 &=& \left[g_{\mu \nu}^{(0)} + \delta g_{\mu \nu} \right] dx^\mu 
dx^\nu 
\nonumber \\
&=& - dt^2+a(t)^2 \left[\gamma_{i j} + h_{i j} \right] dx^i dx^j
\nonumber \\
&=& a(\eta)^2 \left[ - d \eta^2 + \left( \gamma_{i j} + h_{i j} \right) 
dx^i dx^j \right]
\label{metric}
\end{eqnarray}
where $a(t) = e^{H \, t}$ is the scale factor ($H^2 = \Lambda/3$), $\eta$ is the conformal 
time and $\gamma_{i j}$ is the spatially {\em flat} metric (greek indices go 
from $0$ 
to $3$, latin ones from $1$ to $3$ unless otherwise stated). The gravitons 
are traceless, transverse ($h^i_i=0 \,, \partial_j h^{i j} = 0$)
and therefore gauge-invariant with respect to tensorial spatial
transformations. First order scalar metric fluctuations vanish in the
absence of dynamical matter, as already mentioned.

In order to compute the graviton EMT we proceed as in 
textbooks \cite{MTW}:
\be
\tau^{GW}_{\mu \nu} =
- \frac{1}{8 \pi G} G^{(2)}_{\mu \nu} = - M_{\rm pl}^2 \left[ R^{(2)}_{\mu 
\nu}-\frac{1}{2}\left( g_{\mu \nu} g^{\alpha \beta} R_{\alpha \beta} 
\right)^{(2)} \right]
\ee
where we have set $M_{\rm pl}^2 = (8 \pi G)^{-1}$.
The above expression will become, after using the first order equations of 
motion:
\be
\tau^{GW}_{\mu \nu}
= - M_{\rm pl}^2 \left( R^{(2)}_{\mu \nu}-\frac{1}{2} g^{(0)}_{\mu \nu} 
g^{(0) \, \alpha 
\beta} R^{(2)}_{\alpha \beta} \right) \,.
\label{eq_E_tensor}
\ee
where by the superscript $(2)$ we mean terms which are quadratic in the 
perturbation $h_{i j}$. $R^{(2)}_{\mu \nu}$ can be found in Eq. (35.58b) 
of 
\cite{MTW}.

We obtain
\begin{eqnarray}
G^{(2)}_{0 0}&=&-\frac{1}{8}\dot{h}^{i j}\dot{h}_{i j}+\frac{H}{2}
\dot{h}^{i j} h_{i j}+\frac{1}{2}h^{i j}\ddot{h}_{i j}+\frac{3}{8}\partial^k
h^{i j} \partial_k h_{i j} \nonumber \\
& & -\frac{1}{4}\partial^n h^{i m}\partial_m h_{i n}
 \label{G00} \\
G^{(2)}_{i 0} &=& \frac{1}{4}\dot{h}^{m n}\partial_i h_{m n}-\frac{1}{2}
h^{m n}\partial_m \dot{h}_{i n}+h^{m n} \partial_i \dot{h}_{m n}
\label{Gi0} \\
G^{(2)}_{i j} &=& a^2 \delta_{i j}\left[\frac{3}{8}\dot{h}^{m n}\dot{h}_{m n}-
\frac{3}{8}\partial^k h^{m n} \partial_k h_{m n} \right.
\nonumber \\ & & \left.
+\frac{1}{4}
\partial^n h^{i m} \partial_m h_{i n}\right]
+\frac{1}{4}\partial_j h^{m n} \partial_i h_{m n} + 
\frac{a^2}{2} 
\partial^k h^m_j \partial_k h_{m i}
\nonumber \\ &&
-\frac{1}{2} h^{m n} 
\partial_m\partial_j 
h_{i n} 
- \frac{1}{2} h^{m n}\partial_m\partial_i 
h_{j n}+\frac{1}{2} h^{m n}\partial_m\partial_n h_{i j}
\nonumber \\ & &
+\frac{1}{2} h^{m n}\partial_i\partial_j h_{m n}
-\frac{a^2}{2}\dot{h}^m_i \dot{h}_{m 
j}-\frac{1}{2}\partial_m h^n_i 
\partial_n h^m_j  \,,
\label{Gij}
\end{eqnarray}
where $\dot{}$ denotes the derivative with respect to $t$.
The action can also be expanded as $S = S^{(0)} + S^{(2)}$, where 
$S^{(0)}$ is the background value. The second order piece $S^{(2)}$, 
omitting boundary terms, is: 
\ba
S^{(2)} =&& \frac{M_{\rm pl}^2}{8}  \int d^4 x \, a^3  \Biggl[ 
\dot{h}^{m n} \dot{h}_{m n}
- \partial_k h_{m n} \partial^k h^{m n} 
\Biggr] \,.
\label{action}
\ea
Let us perform a Fourier expansion and consider only the physical degrees 
of freedom (polarization states)  $h_+$ and
$h_\times$:
\be 
h_{i j} = \frac{1}{(2 \pi)^3} \int d{\bf k} \, e^{i {\bf k} 
\cdot {\bf x}}
\left[ h_+ e_{i j}^+ + h_\times e_{i j}^\times \right] \,,
\label{polarization}
\ee
where $e^+$ and $e^\times$ are the polarization tensors having the 
following properties ($s =+,\times $):
\be
e_{i j}=e_{j i} \,,\,\,\,k^i \, e_{i j}=0\,,\,\,\,e_{i i}=0\,,
\ee
\be
e_{i j}(-\vec{k}, s)=e_{i j}^{*}(\vec{k}, s)\,,\,\,\,\sum_s
e_{i j}^{*}(\vec{k}, s ) e^{i j}(\vec{k}, s)=4 \,.
\ee
These should be sufficient for our one-loop calculation on shell. Thus we 
do not concern ourselves with unphysical degrees of freedom and ghosts. 
This method of selecting only the physical degrees of freedom was used in 
first deriving the spectrum of gravitational waves  
\cite{staro} and is used in computing the gravitational wave 
contribution to microwave anisotropies.

On quantizing we have 
\be
\hat{h}_{s} (t, {\bf x}) = \frac{1}{(2 \pi)^3} \int d{\bf k}
\left[ h_{s \,, k} (t) \, e^{i {\bf k} \cdot {\bf x}} \,\, \hat{b}_{\bf k} 
+ h_{s \,, k}^* (t) e^{- i {\bf k} \cdot {\bf x}} \,\,
\hat{b}^\dagger_{{\bf k}} \right] \,.
\label{quantumFourier}
\ee
From Eqs. (\ref{action}) and (\ref{polarization}) we see that the 
amplitudes $h_{s \,, k}$ 
satisfy the same equation as massless minimally coupled scalar fields:
\be
\ddot h_{s \,, k} + 3 H \dot{h}_{s \,, k} + \frac{k^2}{a^2} h_{s \,, k} = 
0
\label{eq_motion}
\ee
and the solution for the Fourier mode which becomes a plane wave for short 
wavelengths is:
\be
h_{s \,, k} =  \frac{1}{a^{3/2} \, M_{\rm pl}} \left( \frac{\pi}{2 H} 
\right)^{1/2} H_{3/2}^{(1)}(-k \eta) \,.
\label{ad_vacuum}
\ee
This solution is valid for all values of $k$ except for $k=0$
(corresponding to a zero measure) in which case the solution is simply
a space independent pure gauge.
When averaged over the vacuum state annihilated by $\hat b$ the EMT of 
gravitons takes a perfect fluid form:
\be
\langle \tau_{\mu \nu}^{\rm GW} \rangle = {\rm diag} (\epsilon, a^2 p, 
a^2 p, a^2 p) 
\,,
\ee
which is covariantly conserved in de Sitter 
space-time: $\dot \epsilon + 3 H (\epsilon + p) = 0$ \cite{ABM_PRD,raul}.  

For the vacuum expectation value of the effective energy and pressure
we obtain the following value
\begin{eqnarray}
\epsilon &\equiv& \sum_s M_{\rm pl}^2
\int \frac{d^3 k}{(2 \pi)^3} \, \epsilon_{s \,, k} \nonumber \\
&=& \sum_s M_{\rm pl}^2
\int \frac{d^3 k}{(2 \pi)^3}
\left[ \frac{1}{4} |\dot{h}_{s \,, k}|^2 + \frac{1}{4}\frac{k^2}{a^2}
|h_{s \,, k}|^2 \right.
\nonumber \\
& & \left.
+ H \left( \dot{h}_{s \,, k} h_{s \,, k}^* + 
h_{s \,, k} \dot{h}_{s \,, k}^* \right) 
\right]
\label{bare_energy}
\end{eqnarray}

\begin{eqnarray}
p &\equiv& \sum_s M_{\rm pl}^2
\int \frac{d^3 k}{(2 \pi)^3} \, p_{s \,, k} \nonumber \\
& = & \sum_s M_{\rm pl}^2
\int \frac{d^3 k}{(2 \pi)^3} \left[
-\frac{5}{12} |\dot{h}_{s \,, k}|^2 + \frac{7}{12}\frac{k^2}{a^2}
|h_{s \,, k}|^2
\right] \,,
\label{bare_pressure}
\end{eqnarray}
where $h_{s \,, k}$ are the solutions given in Eq. (\ref{ad_vacuum}).
The above expressions agree with the space averaged EMT of gravitons 
obtained in \cite{ABM_PRD}. It is important to note that the term $H h 
\dot h$ is reminiscent of a scalar field non-minimally coupled to gravity: 
however, a non-minimal scalar field would also have a mass term of order 
$H$ ($\sim H^2 h^2$) which instead is absent in Eqs. 
(\ref{bare_energy},\ref{bare_pressure}). Owing to the presence of the
term $H h \dot h$, the EMT of gravitational waves is not invariant under the
transformation  $h_{ij} \rightarrow h_{ij} + {\rm const.}$ which has
generated so much activity in the context of  
massless minimally coupled scalar fields \cite{KG}. Of course, this is not
surprising at all, since zero modes of gravitons are pure gauge,
as already stated.

In order to regularize bilinear quantities we proceed in the following
way: we subtract the fourth order term of the adiabatic expansion
for the solution to Eq. (\ref{eq_motion}) with a mass term \cite{PF,AP}.
An EMT with a mass term $m$ which regularizes the adiabatic expansion is
needed in order to proceed with the adiabatic subtraction. 
It is then necessary to add a term to the energy density 
($\epsilon_{s \,, k} \rightarrow \epsilon_{s \,, k} (m) = \epsilon_{s \,, k} 
+ m^2 |h_{s \, k}|^2/4$) and to the pressure density 
($p_{s \,, k} \rightarrow p_{s \,, k} (m) = p_{s \,, k} 
+ 5 m^2 |h_{s \, k}|^2/12$) in order to have a
{\em covariantly conserved} EMT corresponding to
(\ref{bare_energy},\ref{bare_pressure}) with a mass term. 
The covariant conservation of such an EMT is given by 
\be
\dot \epsilon_{s \,, k} (m) + 3 H \left[
\epsilon_{s \,, k} (m) + p_{s \,, k} (m) \right] = 0
\ee
and is 
equivalent to the equation of motion (\ref{eq_motion}) with a mass term
$m^2 \, h_{s \, k}$. The regularized energy density is then given by 
\begin{eqnarray}
\epsilon_{\rm REN} = \lim_{k_* \rightarrow \infty}
\lim_{m \rightarrow 0}
\sum_s M_{\rm pl}^2
\int_{|k| < k_*} \frac{d^3 k}{(2 \pi)^3} 
\left[ \epsilon_{s \,, k} - \epsilon_{s \,, k}^{(4)} (m) \right] \,,
\label{ren_energy}
\end{eqnarray}
where $\epsilon_{s \,, k}^{(4)} (m)$ is the fourth order term 
of the adiabatic expansion \cite{BDbook}. Let us note that, 
after performing the integrals, 
we first take the limit $m \rightarrow 0$ and finally remove the 
ultraviolet cut-off: in this manner the correct adiabatic subtraction is 
implemented. Indeed when this technique is
applied to a massless minimally coupled scalar field all the results for
an $E(3)$ invariant-state \cite{AF} are reproduced.
The above technique for computing integrals is therefore
slightly different from the one we previously employed \cite{FMVV,FMVV_2}.

For the EMT we finally obtain 
\be
\langle \tau_{\mu \nu}^{\rm GW} \rangle_{\rm REN} = 
- g_{\mu \nu} \frac{361}{960 \pi^2} H^4 \,.
\label{gw}
\ee
This is our main result, which is in contrast with others claiming that 
gravitons decrease the effective cosmological constant at the one-loop 
order in de Sitter space-time \cite{ford,TW}. 

We stress that the result 
(\ref{gw}) is de Sitter invariant, although the vacuum chosen was
$E(3)$ invariant. The same happens for massless minimally coupled scalar 
fields \cite{BF}. The interesting terms which break de Sitter invariance 
in the regularized value of EMT found in \cite{Folacci,KG} for an $O(4)$ state
in the case of massless minimally coupled scalar fields are due to the use 
of closed spatial sections and are quickly redshifted 
after few e-folds of exponential expansion~\footnote{
Note however that such a result for a massless scalar field
is not valid for physical gravitons, whose spin is $2$. Therefore, the EMT
does not get the de Sitter breaking term due to the
isolated zero modes in a closed spatial section~\cite{ACK}.}. 
We also note that the result in Eq. (\ref{gw}) is obtained from
the integration of the finite terms of the adiabatic expansion: 
this is also what happens for the averaged EMT of
massless minimally coupled scalar fields.

It is important to note that the renormalized EMT of gravitons in
Eq. (\ref{gw}) has $p=-\rho$ as equation of state. The relation
$p=-\rho/3$ is the corresponding unrenormalized one for 
long-wavelength modes \cite{ABM_PRD}.

Let us again focus on the term $2 H h \dot h$: its Fourier component is 
\be
\dot h_{s \,k} h_{s \,k}^* + h_{s \,k} \dot h_{s \,k}^*
= - \frac{H^3 \eta^2}{M_{\rm pl}^2 k} \,,
\label{mixed}
\ee
which is only quadratically divergent in the ultraviolet (in contrast 
with the kinetic and gradient terms) and does not lead
to any bad infrared behaviour (in contrast to $|h_{s \,k}|^2$). The term 
in Eq. (\ref{mixed}) is therefore negative, with a sign which is opposite 
to that of the kinetic and gradient terms:
this difference of sign is also reflected in the renormalized values,
leading to the positivity of the energy density for gravitons.

The result should be compared with that obtained for a massless 
minimally coupled scalar field in the Allen-Folacci (AF henceforth) vacuum 
\cite{AF,BF}:
\be
\langle T_{\mu \nu} \rangle_{\rm REN}^{{\rm AF}} = g_{\mu \nu}
\frac{119}{960 \pi^2} H^4 \,,
\label{desitter_AF}
\ee
which corresponds to a contribution with a negative energy density.
On considering the AF vacuum for the correlator one obtains 
the following result \cite{AF,BF}:
\be
\langle h_s^2 \rangle_{\rm REN} = \frac{H^3 t}{4 \pi^2 M^2_{pl}} \,.
\label{leading_rigid}
\ee
in which de Sitter invariance is broken in the standard way \cite{log}.
When the time derivative of Eq. (\ref{leading_rigid}) is considered,
one obtains
\be
\langle h_s \dot{h}_s \rangle_{\rm REN} =\frac{1}{M^2_{pl}} 
\frac{H^3}{8 \pi^2}  \,,
\label{leading_gw}
\ee
which is the same result as given by our method. 
Thus our approach is consistent with the choice of the AF vacuum.
The reason  for the difference between (\ref{gw}) and (\ref{desitter_AF}) 
is the presence of the term $2 H h \dot h$ for gravitons. Our main result 
(\ref{gw}) can be easily verified by noting that the renormalized 
energy density of gravitons given by Eqs. (\ref{bare_energy}) is
\begin{eqnarray}
\epsilon_{\rm REN} &=& \epsilon_{\rm REN}^{\rm AF} + M^2_{pl} \, 2
H \sum_s \langle h_s \dot{h}_s
 \rangle_{\rm REN} \nonumber \\ 
&=& - \frac{119}{960 \pi^2} H^4 + \frac{H^4}{2 \pi^2} = 
\frac{361}{960 \pi^2} H^4
\,.
\label{easy}
\end{eqnarray}
The contribution of the term $2 H h \dot h$ is positive and larger than 
the (negative) energy density of a massless minimally coupled scalar 
field.

For the case of conformally invariant fields the EMT is independent of  
the vacuum state chosen and is fully given by the trace anomaly $T$:
\be
\langle T_{\mu \nu} \rangle_{\rm REN} = \frac{g_{\mu \nu}}{4}  
T
\ee
with $T$ given by \cite{duff}:
\begin{eqnarray}
T &=& \alpha \, \Box R 
- \frac{\beta}{2} \left( R_{\alpha \beta \gamma \delta} R^{\alpha \beta 
\gamma \delta} - 4 R_{\alpha \beta} R^{\alpha \beta} + R^2 \right)
\nonumber \\
& & + \gamma \, C_{\alpha \beta \gamma \delta} C^{\alpha 
\beta \gamma \delta} \,,
\end{eqnarray}
where $C_{\alpha \beta \gamma \delta}$ is the Weyl tensor, which is zero 
for a metric which is conformal to Minkowsky as is Robertson-Walker (and 
therefore de Sitter).
The coefficients $\alpha$ and $\beta$ obtained by dimensional 
regularizations for scalar, four-component spinors, and gauge fields, 
are respectively \cite{fischetti}:
\begin{displaymath}
\alpha = \frac{1}{2880 \pi^2} \left\{ \begin{array}{l} 1 \\ 6 \\ 12 
\end{array} \right. \,, \quad \beta = \frac{1}{2880 \pi^2} \left\{ 
\begin{array}{l} 1 \\ 11 \\ 62
\end{array} \right..
\end{displaymath}

For (massless) conformally coupled scalar fields one has:
\be
\langle T_{\mu \nu} \rangle_{\rm REN} = - g_{\mu \nu} \frac{H^4}{960 
\pi^2} \,,
\ee
and therefore gravitons contribute, for example, as 361 conformally 
coupled scalar fields!

It is interesting to also investigate the effect of gravitons on the
background space-time within the framework of second order
perturbation theory for the Einstein equations, by evaluating the
gauge invariant (for de Sitter space) geometric quantity $\Theta$
associated with the expansion rate of the universe
(see for example \cite{FMVV_2}).
Hence we consider the following second order metric fluctuations for the
gauge fixed metric given in Eq. (\ref{metric}):
\ba
&& \delta g_{00}^{(2)} = -2 \alpha^{(2)} \,\, , \,\,
\delta g_{0i}^{(2)} = -\frac{a}{2} \beta^{(2)}_{,i} \, , \nonumber \\
&&\delta g_{ij}^{(2)} = \frac{a^2}{2}
\left( \partial_i \chi_j^{(2)}+\partial_j \chi_i^{(2)}+
h_{ij}^{(2)}\right)\, ,
\label{metric2}
\ea
where in the second line the vector $\chi_i^{(2)}$ is divergenceless and the
tensor $h_{ij}^{(2)}$ is transverse and traceless.
The expansion scalar is defined by $\Theta=\nabla_\mu u^\mu$ (where
$u^\mu$ is a normalized vector field, $u_\mu u^\mu=-1$, defining the
comoving frame) and simplifies to
\be
\Theta=3H -\frac{1}{2} h_{ij} \dot{h}^{ij}-3 H \alpha^{(2)} +\frac{1}{a}
\nabla^2 \beta^{(2)} \,.
\ee
On using the Einstein equations one obtains the second
order fluctuations as functions of the physical gravitons.
In particular we find for the expansion
scalar averaged over the vacuum:
\be
\langle \Theta \rangle = 3H \left( 1+ 
\frac{121}{2880 \pi^2} \frac{H^2}{M_{pl}^2} \right) \, .
\ee
Thus we see that the choice of vacuum for the physical gravitons,
which led to the result Eq. (\ref{gw}) corresponding to a
positive cosmological constant contribution, also leads to a
contribution of the same sign to the scalar expansion rate $\Theta$.
 
Our result is only in apparent contradiction with the possibility that scalar 
fluctuations act against the accelerated expansion in chaotic inflation 
\cite{FMVV_2}. A possible explanation of the difference is the
stability of the space-time backgrounds: the inflationary era in
scalar field driven universes is a transient state (local attractor), while
de Sitter is a global solution.

To conclude we have computed the regularized graviton EMT in de Sitter 
space-time by quantizing only the physical degrees of freedom. We have 
found that the (one-loop) graviton contribution to the cosmological 
constant in the $E(3)$ invariant-vacuum is positive, in contrast 
with that of a massless minimally coupled scalar field. 
This effect also appears in a second order perturbative analysis of
the geometrical quantity $\Theta$, which shows an increased expansion rate. 
According to this result, gravitons may then add to the trace anomaly
of conformally 
coupled matter in driving the Starobinsky model of inflation 
\cite{starobinsky}. The 
contribution of gravitons to the cosmological constant is not negligible 
and corresponds to a large number (361) of conformally coupled scalar 
fields. This contribution may also alter 
the inflationary phase of the Starobinsky model 
\cite{starobinsky,vilenkin} since gravitons are 
not conformally coupled, thus the back-reaction may alter the evolution of 
the gravitons themselves. 

Since gravitons are not conformally coupled \cite{grischuck}
the averaged EMT may be state dependent. One may also worry about the 
problems concerning zero modes which plagued 
massless minimally coupled scalar fields \cite{AF,BF,Folacci,KG}. 
Gravitational waves do not have zero modes (insofar as these
correspond to a pure gauge), i. e. $k > 0$.
Further all the contributions from the infrared to renormalized bilinear 
quantities which we compute in this paper are 
finite (we may even include the contribution from the $k=0$ mode
since the measure of this point in Fourier space is zero).

For a massless minimally coupled 
test scalar field in de Sitter space-time, it has been shown that the EMT 
evaluated in the AF vacuum is an asymptotic attractor among all possible vacua 
\cite{AEHMMP}.
If this were true for gravitational waves also, the result obtained for the 
$E(3)$ state would be completely general and lead to an asymptotic value for 
the generalized anomaly~\cite{AEHMMP} $Q^2 = 361/180$
for gravitational waves.

From the theoretical point of view it would also be interesting to see if our 
result (classical and quantum gravity in de Sitter space) can be related 
to conformal field theory, as suggested by the dS/CFT
correspondence~\cite{strominger}.

Last, but not least, should the same result persists for cosmologies with 
$\dot H \ne 0$ and non-vacuum states for modes on large scales, it would 
be interesting to compute the contribution of cosmological perturbations 
to the present energy density.

{\bf Acknowledgments}

\noindent
We would like to thank R. Abramo, B. Allen, L. Ford, B. Losic, K. 
Kirsten, L. Parker, R. Woodard and S. Zerbini for useful comments
and communications.


\begin{thebibliography}{99}

\bibitem{parker} L. Parker, {\em Phys. Rev.} {\bf 183}, 1057 (1969).

\bibitem{BDbook}
N. D. Birrell and P. C. W. Davies, {\em Quantum Fields in Curved Space}
(Cambridge University Press, Cambridge, 1982).

\bibitem{geon}
R. D. Brill and J. B. Hartle, Phys. Rev. {\bf 135}, B271 (1964); 
see J. A. Wheeler, Phys. Rev. 
{\bf 97}, 511 (1955) for the electromagnetic analogue. 

\bibitem{ford}
L. H. Ford, Phys. Rev. {\bf D 31}, 710 (1985).

\bibitem{starobinsky}
A. A. Starobinsky, Phys. Lett. {\bf 91 B}, 99 (1980).

\bibitem{MTW}
C. W. Misner, K. S. Thorne, and J. A. Wheeler, {\em Gravitation}, 
(Freeman, New York, 1973).

\bibitem{staro}
A. A. Starobinsky, JETP Lett. 30, 682 (1979).

\bibitem{ABM_PRD}
L. R. Abramo, R. Brandenberger, and V. F. Mukhanov, {\em Phys. Rev.} {\bf
D 56}, 3248 (1997).

\bibitem{raul}
L. R. Abramo, {\em Phys. Rev.} {\bf D 60}, 064004 (1999).

\bibitem{KG}
K. Kirsten and J. Garriga, {\em Phys. Rev.} {\bf D 48}, 567 (1993).

\bibitem{PF}
L. Parker and S. A. Fulling, {\em Phys. Rev.} {\bf D 9}, 341 (1974).

\bibitem{AP}
P. Anderson and L. Parker, {\em Phys. Rev.} {\bf D 36}, 2963 (1987).

\bibitem{AF}
B. Allen, {\em Phys. Rev.} {\bf D 32}, 3136 (1985);
B. Allen and A. Folacci, {\em Phys. Rev.} {\bf D 35}, 3771 (1987).


\bibitem{FMVV}
F. Finelli, G. Marozzi, G. P. Vacca, and G. Venturi,
{\em Phys. Rev.} {\bf D 65}, 103521 (2002).

\bibitem{FMVV_2} F. Finelli, G. Marozzi, G. P. Vacca, and G. Venturi,
{\em Phys. Rev.} {\bf D 69}, 123508 (2004).


\bibitem{TW}
Note that this is not in contrast with the two-loop effect discussed by 
Tsamis and Woodard, Nucl. Phys. {\bf B 474}, 235 (1996).

\bibitem{BF}
D. Bernard and A. Folacci, {\em Phys. Rev.} {\bf D 34}, 2286 (1986).

\bibitem{Folacci}
A.~Folacci,
J.\ Math.\ Phys.\  {\bf 32}, 2828 (1991) 
[Erratum-ibid.\  {\bf 33}, 1932 (1992)].

\bibitem{ACK}
B.~Allen, R.~Caldwell and S.~Koranda,
Phys.\ Rev.\ D {\bf 51}, 1553 (1995).

\bibitem{log}
A. D. Linde, {\em Phys. Lett.} {\bf 116B}, 335 (1982);
A. A Starobinsky, {\em Phys. Lett.} {\bf 117B}, 175 (1982);
A. Vilenkin and L. Ford, {\em Phys. Rev.} {\bf D 26}, 1231 (1982).

\bibitem{duff}
M. J. Duff, Nucl. Phys. {\bf B 125}, 334 (1977).

\bibitem{fischetti}
M. V. Fischetti, J. B. Hartle, and B. L. Hu, Phys. Rev. {\bf D 20}, 1757
(1979).

\bibitem{vilenkin}
A. Vilenkin, Phys. Rev. {\bf D 32}, 2511 (1985).

\bibitem{grischuck}
L. P. Grishchuk, Zh. Eksp. Teor. Fiz. {\bf 67}, 825 (1974) (Sov. Phys. 
JETP, {\bf 40}, 409 (1975)).

\bibitem{AEHMMP}
P. R. Anderson, W. Eaker, S. Habib, C. Molina-Paris,
and E. Mottola, {\em Phys. Rev.} {\bf D 62}, 124019 (2000).

\bibitem{strominger}
A. Strominger, JHEP 0110, 034 (2001).

\end{thebibliography}
\end{document}